\documentclass{article}
\usepackage{graphicx}

\begin{document}
\centerline{LAWS, SYMMETRIES, AND REALITY\footnote{This paper is based on the invited talk given by the author at the Wigner Centennial Conference,
Pecs, Hungary, 8-12 July, 2002.}}
\vskip 1cm
\centerline{Jeeva Anandan}
\centerline{Department of Physics and Astronomy}
\centerline{University of 
South
Carolina}
\centerline{Columbia, SC 29208, USA.}

\centerline{and}
\centerline{Clarendon Laboratory}
\centerline{University of Oxford}
\centerline{Parks Road}
\centerline{Oxford, ~~UK}

\centerline{E-mail: jeeva@sc.edu}

\begin{abstract}

 It is argued that quantum mechanics follows naturally from the 
assumptions that there are no fundamental causal laws but only 
probabilities for physical processes that are constrained by 
symmetries, and reality is relational in the sense that an object is 
real only in relation to another object that it is interacting with.   
The first assumption makes it natural to include in the action for a 
gauge theory all terms that are allowed by the symmetries, 
enabling cancellation of infinities, with 
only the terms in the standard model observable at the 
energies at which we presently do our experiments. In this approach, it is also 
natural to have an infinite number of fundamental interactions.

\end{abstract}
\bigskip
\centerline{April 14, 2003}

\newpage

\section{Introduction}

We are told that nature is fundamentally `capricious'. For example, 
if we prepare many identical atoms in the {\it same} excited state at 
a given time, then they decay at different times, in general. This  has 
been a source  of tremendous surprise, if not  shock,  for many 
physicists who were brought up to accept, without question, the 
paradigm of laws \cite{an1999}. They believe that all physical 
systems are governed by the same laws, which should compel 
identical systems having the same initial conditions to behave in the 
same way. But if we are unbiased by centuries of traditional physics 
conditioning, then we should actually be surprised if all the atoms 
behave in the same way. Why should they? After all, they do not 
communicate with each other to ensure that they would all decay at 
the same time. And there is no evidence of a `cosmic rail road,'  i.e. 
fundamental laws of nature, that would compel all the atoms to 
behave in the same way \cite{an2002}.

The view that there are fundamental laws worked very well in 
classical physics until it was found that the laws of classical physics 
are really not fundamental. When physicists found that the laws of 
classical physics were in disagreement with observation, they 
decided to replace these laws with new laws. Although they 
discarded the existing laws, they did not give up the belief that there 
must be laws. Newton's second law of motion was replaced by 
Schrodinger's law that was supposed to govern the evolution of a 
quantum state. While this law is supposed to govern the {\it 
unobserved} evolution of the state between measurements, it does 
not appear to govern the {\it observation} of a quantum state, i.e. 
the process of measurement. 

Some physicists found this state of affairs to be highly 
unsatisfactory. They wanted the quantum laws to apply a to every 
physical process including the process of measurement. Everett 
\cite{ev1957}, for example, postulated that the Schr\"odinger law 
applied to every quantum evolution, which implied that the wave 
function never collapsed.  Bohm \cite{bo1952}, on the other hand, 
regarded the wave function and the particle to be real, the wave 
function playing the dual role of giving the probability density for 
the particle and guiding the particle's motion according to a 
quantum law.  Another approach, pioneered by Pearle 
\cite{pe1986}, was to modify Schr\"odinger's law to a new non 
linear law that would apply to the measurement process.

These and similar approaches to quantum theory were based on two 
assumptions: A) There are quantum laws that apply to every 
physical process, including the measurement process, and B) a 
system may exist by itself and its reality does not depend on its 
interaction with other systems, which I shall call the assumption of 
absolute reality. But the `capriciousness' of nature mentioned at the 
beginning suggests that the assumption (A) may not be valid. The 
great difficulty in applying quantum laws to the observed 
measurement process suggests that these laws may not apply also to 
the unobserved state evolution in between measurements. The 
strangeness of assumption (B) is seen if we imagine a universe 
consisting of only one object. What is the metaphysical difference 
between this object existing and not existing? But if there are two 
objects then these two objects may interact, and each object would 
then have reality with respect to the other.

 It seems reasonable therefore to suppose, instead, that 1) there are 
no fundamental causal laws but only probabilities for physical 
processes that are constrained by symmetries, and 2) reality is 
relational in that an object is real only in relation to another object 
that it is interacting with. These two assumptions help us to 
understand why the world is
quantum mechanical. Assumption (1) implies that nature must 
necessarily be indeterministic or `capricious,' which is consistent 
with observed quantum phenomena. Assumption (2) explains why 
the act of measurement brings into being the state of a quantum 
system.  This is consistent with the Copenhagen interpretation that 
denies absolute reality. But the present interpretation goes 
beyond the Copenhagen interpretation by replacing the absolute 
reality with relational reality. This allows for an objective reality that 
is relational. According to the present view, a tree falls in a forest 
even when there is no one to observe it, because of the interactions 
between the molecules constituting the tree and their interaction 
with the environment.  

A version of relational reality was proposed by Rovelli \cite{ro1996}
and Mermin \cite{me1998}, who argued for the reality of 
``correlations without correlata.'' But this interpretation of 
quantum mechanics seems to be indistinguishable from the Everett 
interpretation, which also has all the correlations in the wave 
function of the universe.  The present interpretation differs from 
these interpretations in two ways: First the relational reality is associated 
with the {\it interactions} and not correlations. For example, the EPR 
correlation between two non-interacting spin-half particles is due to 
an interaction that the two particles have undergone in the past.  
Therefore, the measurements on an ensemble of such pairs that 
confirm these correlations are due to the earlier interaction that the 
particles underwent. Second, the implication of assumption (1) that the world is indeterministic introduces probabilities from the very beginning. Whereas, obtaining quantum probabilities in the deterministic picture of Everett is a major problem.

In section \ref{mysteries}, I shall discuss the fundamental role played by the 
Poincare group of symmetries in quantum mechanics, and argue that 
symmetries are more basic than laws.  This argument will be extended, in section \ref{gauge}, to include gauge symmetries. I shall then show, in section \ref{quantum}, 
that this point of view naturally leads to quantum mechanics. 
Relational reality will be used to justify the Born rule for obtaining quantum probabilities. 
Finally, in section \ref{beyond}, I shall argue, on the basis of assumption 
(1), that there should not be a finite symmetry group  that gives all the fundamental interactions. 
This suggests that there must be an infinite number of gauge 
interactions associated with the groups $SU(N), N=1,2,3....$.

It is a great pleasure to dedicate this paper to Eugene Wigner who, along with  Hermann Weyl, pioneered  the use of symmetries in quantum theory, and  had a deep understanding  and appreciation of the profound role of symmetries in physics \cite{wi1970}.

\section{Mysteries and Symmetries in Quantum Mechanics}
\label{mysteries}

One of the most mysterious aspects of quantum mechanics is the 
{\it wave-particle duality}. The state of the system may be in an 
approximate eigenstate of momentum $p$, in which case it may be 
regarded as a wave, or in an approximate eigenstate of position 
$x$, in which case it may be regarded as a particle.   The wave 
particle duality is therefore due to $x$ and $p$ being independent 
observables in quantum theory. Closely related to this aspect is 
{\it complementarity} that is implied by the Heisenberg commutator 
the relation: 
\begin{equation}
[x_j,p_k]=i\hbar\delta_{jk},~~j,k=1,2,3
\label{heisenberg}
\end{equation}
This is unlike in Newtonian physics where the momentum is defined 
by $p=m{dx\over dt}$, and therefore $x$ and $p$ are not 
independent.  Moreover, $x$ and $p$ commute in classical physics.

 Both aspects in quantum theory may be understood by realizing 
that $x$ and $p$ are independent generators of the inhomogeneous 
Galilei group that is the symmetry group of non relativistic quantum 
mechanics. $p$  generates spatial translation and $x$  generates 
Galilei boost, and therefore they are independent, which gives rise 
to the wave-particle duality. During a measurement, what is observed 
is the {\it relation} between the apparatus and the observed 
system. This relation therefore should be regarded as the 
observable. This is the fundamental reason why observables are 
operators in quantum theory. These relations are elements of the 
symmetry group that constitutes the quantum geometry 
\cite{an1999}.

To understand the complementarity between $x$ and $p$, consider 
the Poincare Lie algebra relations:
\begin{equation}
[K_j,T_k]=i\delta_{ik}T_0, ~~j,k=1,2,3,~~[T_\mu,T_\nu]=0
\label{poincare}
\end{equation}
 where $ iK_j$ generate Lorentz boosts and $iT_\mu,~\mu=0,1,2,3$
generate space-time translations. Here $ K_j$ are dimensionless 
because they get multiplied by the dimensionless parameters $v/c$ 
and exponentiated to obtain the infinitesimal Lorentz boosts, and 
$T_\mu$ have the dimension of $1/$length because they get 
multiplied by distances and exponentiated to get the translations, 
and the exponents of course must be dimensionless. In order to 
relate $T_j$ to the momentum $p_k$ that is conserved under 
translation, it is therefore necessary to introduce a new scale that 
has the dimension of momentum $\times$ length. This new 
fundamental constant, denoted $\hbar$,  enables also the time 
translation $T_0$ to be related to the energy $p_0$ that is 
conserved due to time translational symmetry. We then write
\begin{equation}
\hbar T_j = p_j, ~j=1,2,3, ~~\hbar c T_0=p_0
\label{energy-momentum}
\end{equation}
The introduction of Planck's constant here seems to be related to 
the space-time description in physics.

The Lorentz transformations are generated by $L^{\mu\nu}=x^\mu 
T^\nu- x^\nu T^\mu $. Then 
\begin{equation} 
K_j =L^{j0}=x^j T_0 + x^0 T_j .
\label{boost}
\end{equation} 
Substituting this in the first relation in (\ref{poincare}), and using 
(\ref{energy-momentum}),
\begin{equation}
[x^j,p_k]p_0 = i\hbar \delta_{ij}p_0, ~~j,k=1,2,3
\label{preheisenberg}
\end{equation}
Now $\eta_{\mu\nu}T^\mu T^\nu =\hbar^{-2}c^{-2}
( {p_0}^2-{\bf p}^2)$ is a Casimir operator that commutes with the 
Poincare group. For a given irreducible representation, we may 
therefore set
$ {p_0}^2-{\bf p}^2= m^2c^4$,  where $m$ is the mass.
 This implies that at low energies, $ p_0\approx mc^2$. Then 
(\ref{preheisenberg}) becomes (\ref{heisenberg}). We could have 
obtained (\ref{heisenberg})by writing 
$T_\mu=i{\partial\over\partial x^\mu}$ and using 
(\ref{energy-momentum}). But the purpose of the above exercise is to show that 
(\ref{heisenberg}), which is so fundamental to quantum mechanics, 
and the Heisenberg uncertainty principle of that follows from it, 
ultimately come from the Poincare Lie algebra relations 
(\ref{poincare}).

Consider now the Poincare Lie algebra relation
\begin{equation}
[K_j,T_0]=iT_j, ~~j=1,2,3.
\label{poincare2}
\end{equation}
 On using (\ref{boost}) and (\ref{energy-momentum}) this reads
\begin{equation}
{1\over c^2}[x^j,p_0]p_0=i\hbar p_j, ~~j=1,2,3.
\label{poincare3}
\end{equation}
Writing $p_0=mc^2+H$, at low energies
\begin{equation}
[x^j,H]m=i\hbar p_j, ~~j=1,2,3.
\label{poincare4}
\end{equation}
 This is satisfied by $H={1\over 2m}{\bf p}^2 + V$, where $V$ 
commutes with $x^j$. It follows also from the Poincare Lie algebra 
relations that $[p_j,H]=0$ and $[J_k,H]=0$,  where $J_k$ generate 
rotations. However, the above assumption of Poincare symmetries 
implies that this non relativistic Hamiltonian $H$ corresponds to an 
isolated system that is non interacting.

The above procedure leads to the In\"on\"u-Wigner contraction 
\cite{in1952} of the Poincare group to the quantum mechanical 
inhomogeneous Galilei group.  It was shown by Jauch \cite{ja1968} that the 
covariance of time evolution under the homogeneous Galilei group 
requires that the Hamiltonian that generates time evolution must be 
of the form
\begin{equation}
H={1\over 2m}[ {\bf p}-{\bf A}({\bf x},t)]^2 + V({\bf x},t)
\label{hamiltonian}
\end{equation}
Thus symmetries determine the `law' of quantum evolution as well 
as the interaction of the quantum state that includes the 
electromagnetic interaction.

 Another important consequence of the Poincare symmetries is seen 
by taking the expectation value of with respect to the vacuum of the 
first of the relations (\ref{poincare}) with $j=k$:
\begin{equation}
<0|[K_j,T_j]|0>=i <0|T_0|0>, 
\label{vacuum}
\end{equation}
 Since the vacuum is invariant under translations or boosts, the left 
hand side is zero.  Therefore, $<0|T_0|0>=0$, which implies that the 
vacuum energy is zero.  But, as is well known, if we apply the laws of 
quantum mechanics to fields then the vacuum energy is {\it infinite} due 
to the fact that the fields consist of infinite number of harmonic 
oscillators that have zero point of energies. This shows that the 
symmetries should be regarded as more fundamental then the 
`laws' that are applied to these harmonic oscillators.

It is now tempting to say that the above fundamental role 
played by the Poincare and Galilei groups in giving rise to 
quantum mechanics is due to the existence of space-time on 
which these groups act as symmetries.  But the experimentally 
observed intrinsic spin suggests that the symmetries are more 
fundamental then the space-time description. The generators 
$J_k$ of the rotation group $SO(3)$ that is a subgroup of the 
Poincare group are the components of angular momentum. The 
commutator relations of these components are therefore the 
same as the Lie algebra relations of the rotation group.  The 
intrinsic spin that is contained in these generators cannot 
be obtained from the space-time description that gives only 
the orbital angular momentum. Also, as is well known, in this 
way only the integer spin particles, or Bosons, are obtained. 
To obtain half integer spin particles, or Fermions, it is 
necessary to postulate $SU(2)$ that is the covering group of 
$SO(3)$ as the symmetry. We cannot regard a Fermionic wave 
function as `immersed' in space time because when it is 
rotated by $2 \pi$ radians it changes sign, which has 
observable consequences \cite{ah1967}.  The $SU(2)$ group 
which contains this transformation cannot therefore be 
associated with the symmetries of space.  Also, the addition 
rules for angular momenta of two systems can only be 
understood by considering irreducible representations of 
$SU(2)$ group in the tensor product of the Hilbert spaces of 
the two systems, and not by regarding angular momentum vector 
as representing rotation of matter in space about an axis 
with arbitrary direction. To obtain this group, the usual 
space-time Poincare group $P$ needs to be replaced by the 
semi-direct product $\tilde P$ of $SL(2,C)$ and the 
translational group.  And $SU(2)$ is a subgroup of $SL(2,C)$. 
Since $\tilde P$ is not a direct consequence of the usual 
space-time description, we should regard $\tilde P$ as being 
more fundamental than space-time.

\section{Structure of a Gauge Theory}
\label{gauge}

More generally, in relativistic quantum theory, the quantum system 
interacts with a general gauge field. The gauge symmetry group 
implies, via Noether's theorem, conserved quantities. These 
conserved quantities generate fields, which is believed to be in accordance 
with a `law'.  For example, in electromagnetism the gauge symmetry 
group is  $U(1)$ and the conserved quantity is the electric charge  
that generates the electromagnetic field in accordance with 
Maxwell's law.  The great achievement of Weyl and Yang-Mills was to 
recognize that the third side of this triangle 
(Fig. 1) may be completed by 
means of the gauge principle, i.e. the gauge fields may be obtained 
directly from the gauge group by requiring local gauge symmetry. 

\begin{figure}
\begin{center}
\includegraphics [width=3in]{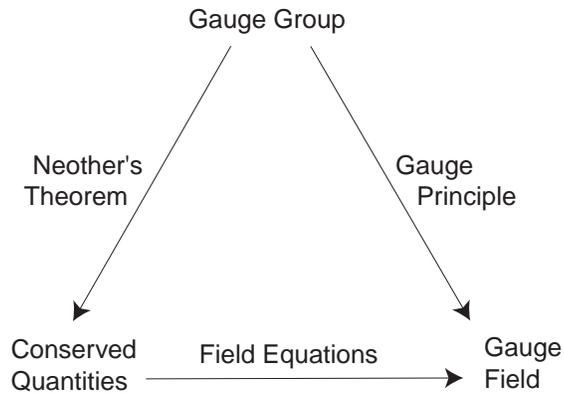} 
\caption{ The structure of a gauge theory.  The 
fundamental role played by the gauge symmetry group is shown 
from the fact that the gauge field may be obtained from the gauge 
group via the conserved quantities or directly by means the gauge 
principle. The requirement that both paths are equivalent suggests that all possible invariants may play a role in the Lagrangian for the gauge field.}
\end{center}
\end{figure}
The first side of this triangle, namely Noether's theorem that gives 
the conserved quantities from the symmetries is tautological 
\cite{an2002b} as mathematical theorems are.  The third side may 
be made nearly tautological, if the gauge principle is implemented 
not in a particular Lagrangian but as follows: In order to compare 
vectors that belong to internal spaces at different space-time points 
it is necessary to introduce a connection that enables parallel 
transport of a vector from one point to another. In fact, the gauge 
field may be introduced by giving the holonomy transformations as 
elements of the symmetry group from which the connection may be 
reconstructed, and the connection is then unique up to gauge 
transformations \cite{an1983}. This shows that gauge fields may be 
obtained directly from symmetries without a dynamical or causal 
law. 
On the other hand, for the second side, it is usually supposed that 
the conserved quantity rigidly determines the gauge field according 
to a `law'. This would make the second side fundamentally different from 
the first and the third sides.

But I shall suppose now that going from the first vertex (symmetries) to 
the third vertex (gauge field) along the third side is in some sense 
equivalent to going along the first and the second sides. This   
implies that the second side should not pick out a particular form of the
Lagrangian or Hamiltonian like the first and the third sides. 
Therefore, we should {\it not} suppose that the Lagrangian for the gauge 
field is the Yang-Mills Lagrangian that is proportional to the Lorentz-
gauge invariant
\begin{equation}
I_1 = {F^i}_{\mu\nu} {F^i}^{\mu\nu}
\label{ Yang-Mills}
\end{equation}
 For the $SU(2)$ gauge field, for example, any  Lorentz-gauge 
invariant that is a polynomial in the field strength $ {F^i}_{\mu\nu} $ 
is a polynomial function of ten Lorentz gauge invariants that are polynomials 
in ${ F^i}_{\mu\nu} $, of which only nine are independent.  The 
simplest of these invariants is 
(\ref{ Yang-Mills}). But there are other 
invariants such as $I_2=\epsilon_{ijk} {F^i}_{\mu\nu} 
{F^j}^{\nu\rho} {{F^k}_\rho}^\mu$ or $I_3={F^i}_{\mu\nu} 
{F^j}^{\mu\nu} {F^j}_{\rho\sigma} {F^i}^{\rho\sigma} $ on which 
the Lagrangian may depend on. I shall therefore suppose, in 
accordance with the above hypothesis, that the Lagrangian depends 
on all ten invariants. More generally, for an arbitrary gauge theory, I 
shall assume that the Lagrangian is a function of all the Lorentz-gauge invariants.

This makes all three sides of the triangle in Fig. 1 
similar in the sense that none of them 
picks out a particular invariant to determine the dynamics. 
However, I do not have an explanation for the values of coupling 
constants that should appear in the Lagrangian apart from appealing 
to a version of the anthropic principle: There are parallel universes 
that constitute what I have called polyverse \cite{an2002}, and it 
then follows that we must live in a universe with coupling constants 
that enable life to evolve.

According to the standard model, three of the four 
fundamental interactions that are known today are described 
by a gauge theory with the gauge group $U(1)\times 
SU(2)\times SU(3)$.  What about the remaining interaction, 
namely gravitation? From an experimental point of view, the 
best `law' that we have for gravitation are the classical 
Einstein's field equations. But this `law' does not determine 
the signature of the metric. The Lorentzian signature of the 
metric, however, is obtained by requiring invariance of the 
metric at each point under the Lorentz group of symmetries. 
This suggests that for gravity as well symmetries maybe more 
fundamental than laws. Indeed, it is possible to characterize 
the gravitational field by associating an element of the 
Poincare group with each (piece-wise differentiable) 
curve on space-time \cite{an1996} analogous to characterizing 
a gauge field by associating gauge group elements with each 
such curve \cite{ya1974}. 

However, hitherto it has been held that a fundamental difference 
between gravity and gauge fields is due to the Lagrangian for gauge 
fields being quadratic in the curvature or field strength whereas the 
Einstein-Hilbert Lagrangian for gravity is linear in the curvature. But 
the above hypothesis that the two paths to gauge field from the 
gauge group should be equivalent removes this fundamental 
distinction.  This is because, according to this hypothesis, all the 
invariants should be included in the Lagrangian for both gravity and 
gauge fields.

It is well known that the Einstein-Hilbert Lagrangian density 
for gravity $\sqrt{-g}R$, 
where  $R$ is the Rcci scalar, is
not renormalizable If one requires that there should be a `law' given 
by a Lagrangian that depends on a small finite number of invariants 
then it becomes necessary to deal with the infinities that arise in 
Feynman amplitudes by the process of renormalization. But if the 
Lagrangian depends on all possible invariants then there are counter 
terms to cancel all the infinities \cite{we1995}. The usual process of 
renormalization that absorbs the infinities in a finite number of 
coupling constants was conceived within the paradigm of laws 
because of the belief that the Lagrangian that constitutes the `law' 
should depend on a small subset of all possible invariants.  But if we 
allow all possible invariants then the infinities may all be cancelled.

The present approach then appears to solve the riddle that arises 
within the paradigm of laws of why nature should
choose particular Lagrangians and not others for the laws.
This is because, according to the present view, all
lagrangians consistent with a given set of symmetries are
allowed. However, in the present approach, symmetries replace the 
fundamental role previously played by laws. And the `laws' are 
obtained from the symmetries, as effective laws, instead of the other 
way around.

\section{ Why the World is Quantum Mechanical}
\label{quantum}

It was argued, in section \ref{mysteries},  that a great deal of 
quantum mechanics may be obtained from the Poincare group of 
symmetries. In section \ref{gauge}, this argument was extended 
to include the gauge symmetries that are used today to describe 
three of the four known fundamental interactions. From an operational point of view, observing the gauge field by a quantum mechanical probe is the same as observing the holonomy transformations that are elements of the gauge group \cite{an1996}.
Hence, symmetries are directly observable in quantum mechanics. But another 
important ingredient of quantum mechanics is the linearity of the 
time evolution of a quantum state, which leads to the quantum 
measurement problem. 
Why should the Hamiltonian obtained using symmetries in section \ref{mysteries}, for example, generate linear time evolutions of 
state vectors? It will now be argued that assumption (1), 
stated in the introduction, naturally leads to
 this linear time evolution, or Schr\"odinger's equation.

Since, according to assumption 
(1), there are no fundamental causal laws, all the infinite possible 
ways in which a system may go from an initial state to a final state 
should have equal probabilities. For this to make mathematical 
sense {\it in the absence of laws}, there should be cancellation 
between the different paths in spite of them having equal 
probabilities \cite{an2002}. This becomes possible only by 
introducing the probability amplitude associated with each path so 
that in order to determine the probability of a process the 
amplitudes should be added for the different ways in which the 
process can take place and the probability is determined from this 
sum. Mathematical considerations then suggest that the probability 
amplitudes should be complex numbers \cite{an2002}. 

The 
requirement that these probability amplitudes should be invariant 
under the symmetries, in accordance with assumption (1), then 
gives quantum mechanics in the Feynman path integral formulation, 
except that the action that is the phase of the probability amplitude 
needs to include all possible terms that are invariant under the 
symmetries. Since the Feynman path integral formulation is 
equivalent to the Schr\"odinger formulation, we obtain the linear time 
evolution of the state vector. The natural origin of the action as the phase of the probability amplitude, in the present approach, provides a possible explanation for the fundamental role played by the action in physics. 

Also, including all possible invariants in the action of a gauge theory 
provides counter terms to cancel all the infinities that arise in 
summing the amplitudes in the quantum field theory, as mentioned in 
section \ref{gauge}. But in order to have consistency with experiments, 
only the lowest order terms in the action that are in the standard model should be observable at the energies at which we do experiments at present.  It can be argued to that such a theory would not have unitary time evolution at large energies. But this argument is based on our present understanding of quantum field theory, which may have to be modified at higher energies.

{\it How} and {\it when} do we convert probability amplitudes into 
probabilities? Quantum mechanics provides a clear answer to the 
question of `how', namely the Born rule, but is infamously 
ambiguous about the question of `when'. We are told that 
probability amplitudes should be added or multiplied when no 
`observation' is made, and that the probability amplitudes should be 
converted to probabilities when an `observation' is made. But no 
clear criteria for what constitutes an observation is given, apart 
from some vague ideas about interaction with a macroscopic 
system. It is my contention that the answer to the above question of 
`when' provides also the answer to the question of `how,' i.e. a 
derivation of the Born rule.

 In classical physics, the events, for which we can only predict 
probabilities in quantum physics, have no ontological ambiguity. 
These events are assumed to exist independently of any observation 
of them. But even in classical physics, events are due to interaction between 
two or more objects, such as the Einsteinian example of lightning 
striking a railway track. The belief in the existence of an object 
independently of its interaction, which I call absolute reality, cannot 
be operationally confirmed. Absolute reality is therefore a 
metaphysical assumption, which cannot even be philosophically 
defined because there does not exist a philosophical criterion for 
distinguishing between `existence' and `nonexistence' other than 
observation.
The assumption of absolute reality may be justified if it's consequences are confirmed 
by observation. But the observer dependence of the quantum state 
suggests that this assumption is not valid.

I shall therefore assume, instead, that reality is relational in the 
sense that two objects exist in relation to each other if and only if 
they interact. How can we speak of objects whose very existence is 
conditional upon their interaction? The statement that `there are no 
ghosts' does not presuppose the existence of ghosts. So, there 
is no contradiction in referring to objects that  do not exist, although in
the present case they would exist in a relational sense when they interact,
which necessitates referring to them. Even in classical physics, the reality
of the electric field is determined by what it does to a charge; the field
is therefore real with respect to the charge that it interacts with. The difference
between classical and quantum physics is that different charges respond to a classical 
field like as if it is the same field, which gives the illusion 
that the field is independent of its interaction with the charge. Whereas, in quantum physics the states of two interacting systems become entangled, in general, which should prevent us from assigning independent reality to either state.

Relational reality leads to the experimentally observed Born rule for 
obtaining probabilities from probability amplitudes, which may therefore be 
regarded as evidence of relational reality. This is most easily seen 
in the double slit interference experiment. If the 
particle going through one of the slits interacts with another 
physical system, then it is this interaction that brings into reality the 
particle going through that slit relative to the physical system that it 
interacts with. If this interaction does not take place then the only 
way the particle could go through the screen with the double slit is 
by passing through the other slit. Therefore, for this arrangement, 
the probability of the particle interacting with any part of the 
subsequent screen is the sum of the two probabilities for passing 
through each of the two slits. This requirement naturally leads to 
the Born rule that this probability is $|\psi|^2$ where $\psi$  is the 
sum of the two probability amplitudes for the particle to go through 
the two slits \cite{an2002}. 

A special case of an interaction is when a human being, or more generally an animal being, makes an observation. What is observed then acquires relational reality with respect to this being. We may therefore understand now Wigner's hypothesis that the `consciousness' collapses the wave function \cite{wi1961} as a special case of relational reality. However, it would be a mistake to assume, as Wigner may have done implicitly, that the collapsed wave function has absolute reality. For example, the Schr\"odinger cat inside a box may be alive in the sense that different parts of its body are accordingly interacting with each other and with the box. But an observer outside the box may {\it expect} the cat to be in a superposition of alive and dead states, on the basis of prior measurements, although there is no interaction to verify this expectation due to the large number of degrees of freedom of the cat. There is no contradiction between the two views because reality is relational. However, the cat acquires relational reality only as alive or dead because of the restrictions on the interactions that it could have with another object. But if the cat is replaced by a microscopic system, then the outside observer may observe this system in the expected superposition by means of a suitable interaction, which is now possible because of its small number of degrees of freedom. This provides a resolution of the Schr\"odinger cat paradox \cite{an2002}.

Descartes' famous statement ``I think, therefore I am'' ({\it Cogito ergo sum}) created, despite Descartes' healthy skepticism of reality, a great deal of confusion in Western philosophy. ``I think'' means interactions between different parts of the brain, which therefore have relational reality with respect to each other. But concluding from this ``I am,'' implying absolute reality of the self, is an unjustified extrapolation. 

\section{Beyond Symmetries}
\label{beyond}

In order to do physics we must communicate information. This 
naturally leads to gauge fields and gravitation \cite{an1986}. The 
symmetry group of the experimentally very successful standard 
model is the direct product of the Poincare group and the gauge 
group. This suggests that the gauge group may also just be a direct 
product of groups, as it already is in the standard model. But if 
there is one finite dimensional symmetry group that determines all 
interactions then this would constitute a law. It may well be that the 
way we observe the universe with the very limited apparati that we 
have which makes symmetries so useful. At the low energies in 
which we do our experiments, the symmetry groups $U(1),SU(2), 
SU(3)$, which are the simplest unitary groups, useful. But at higher 
energies, we may find the gauge groups. $SU(4), SU(5), SU(6),....$  
useful. 

The gauge group $SU(4)$ was used by Pati and Salam \cite{pa1973} 
to unify quarks and leptons by putting the three color states of a 
quark and the corresponding lepton in the same multiplet on which 
the fundamental representation of $SU(4)$ acts.  Also, the smallest 
simple group that contains the gauge group of the standard 
model is $SU(5)$, which has therefore been used in an attempt to 
unify the electroweak and strong interactions \cite{ge1974}.  A 
major problem in such grand unification models is that proton 
decay has not been observed yet. Also, all the grand unification 
attempts implicitly assume that there exists a finite dimensional 
symmetry group that unifies the fundamental interactions. But such 
a symmetry group would constitute a law, and is therefore contrary 
to the spirit of the present paper according to which there is no 
intelligent design to the universe. This suggests that there should be an infinite hierarchy of 
fundamental interactions associated with $ SU(n),n=1,2,3,4,....$

 The universe on a large scale is held together by the gravitational 
interaction,  which is associated with the Poincare group 
\cite{an1983}. If we probe deeper on the scale of molecules and 
atoms, we find that they are held together by electromagnetic 
interactions, corresponding to the $U(1)_{EM}$ symmetry group. 
Owing to the success of the standard model, we should say that the 
electrons and the nucleus in an atom are held together by the 
electro-weak field corresponding to $U(1)\times SU(2)$, 
which leads to the experimentally observed parity violation in atoms.  
If we probe deeper, we observe the strong interactions 
that hold the quarks together in neutrons, protons and other 
hadrons,
associated with the group $SU(3)$. The fact that these are the 
simplest unitary groups suggests that this may be due to the low 
energies of the experiments that we have been doing so far. 
Extrapolating, it would appear that quarks and leptons have 
constituents that are held together by a gauge field of the $SU(4)$ 
group.
But since quarks and leptons have spin$1/2$, we then expect this 
symmetry group to be broken so that 
three of the four particles in the fundamental representation form the 
quarks and leptons. This is analogous to how in the Pati-Salam model the $SU(4)$ symmetry is broken so that the hadrons are formed by the quarks, while the leptons stand apart.

Present experiments have placed an upper limit for the radii of quarks and leptons of about $10^{-17}$ cm. The next generation of experiments in the Lepton-Hadron Collider is expected to probe scales less than $10^{-18}$ cm. So, there is hope that the above extrapolation to a super-strong force of the $SU(4)$  gauge field may be experimentally testable.

 To conclude, we recall Einstein's famous statement that 
our theories are to the external
world what clothes are to the human body. The physical theories 
proposed so far have all been based on the assumption that there 
are fundamental laws. This would mean that
these laws or ``clothes'' may be made to fit the objective
reality more and more closely, but there is an {\it
unbridgeable gap} between them. The purpose of this paper was to 
argue
that these `laws' are like `the emperor's new clothes'. For the 
relational reality obtained through our observations, laws and, at a 
deeper level, symmetries are useful. But neither may be a reflection 
of any fundamental structural realism. 

\bigskip

\noindent{\bf Acknowledgments}

I thank Yakir Aharonov, Pawel Mazur, and
Parameswaran Nair
 for useful discussions. This research was partially
supported by an NSF grant, an ONR grant, and a Fulbright award.


\begin{thebibliography}{99}

\bibitem{an1999}
J. Anandan, Foundations of Physics, {\bf 29,} no. 11, 1647-
1672 (1999), quant-ph/9808045.

\bibitem{an2002}
J. Anandan, Foundations of Physics Letters, {\bf 15,} no. 5, 415-438 
(2002), physics/0112020.

\bibitem{ev1957}
H. Everett , Rev. Mod. Phys., {\bf 29,} 454-462 (1957).

\bibitem{bo1952}
D. Bohm, Phys. Rev., 
{\bf 85,} 166-193 (1952) .

\bibitem{pe1986}
P. Pearle in {\it Quantum Concepts in Space and Time,} eds. 
R. Penrose 
and C. J. Isham (Clarendon Press, 1986), 84-108;  
P. Pearle, Phys. Rev. 
A {\bf 39,} 2277-2289 (1989).

\bibitem{ro1996}
Carlo Rovelli, Int. J. of Theor. Phys. {\bf 35,} 1637 (1996), quant-
ph/9609002.

\bibitem{me1998}
N. David Mermin, Pramana {\bf51,} 549-565 (1998), quant-
ph/9609013; American Journal of Physics 66, 753-767 (1998), 
quant-ph/9801057.

\bibitem{wi1970}
{\it Symmetries and Reflections, Scientific Essays of Eugene P. Wigner} (The M.I.T. Press, Cambridge, MA 1970).

\bibitem{in1952}
E. In\"on\"u and E. Wigner, Nuov. Cim. {\bf IX,} 705 (1952).

\bibitem{ja1968}
J.M. Jauch, {\it Foundations of Quantum Mechanics} (Addison-
Wesley, Reading, MA, 1968).

\bibitem{ah1967}
Y. Aharonov and L. Susskind, Phys. Rev. {\bf 158,} 1237-9 (1967).

\bibitem{an2002b}
J. Anandan, International Journal of Theoretical Physics, {\bf 41,} 
No. 2, 199-220 (2002), quant-ph/0012011.

\bibitem{an1983}
J. Anandan, {\it Conference on Differential Geometric Methods in Theoretical Physics}, Trieste, July 
1981, edited by G. Denardo and H.D. Doebner (World Scientific, 1983), 211-215; J. 
Anandan,  Phys. Rev. D,  {\bf 33, } No. 8, 2280-2287 (1986),

\bibitem{an1978}
J. Anandan,  Journal of Mathematical Physics 19, 260-268 (1978); J. Anandan and R. 
Roskies, Journal of Mathematical Physics 19, 2614-2618 (1978).

\bibitem{an1996}
J. Anandan, Phys. Rev. D  53, No. 2, 779-786 (1996).

\bibitem{ya1974}
C.N. Yang, Phys. Rev. Lett. {\bf 33,} 445 (1974).

\bibitem{we1995}
Steven Weinberg, {\it The Quantum Theory of Fields I} (Cambridge
University Press, Cambridge 1995), pp.144, 145, 198.

\bibitem{an1986}
J. Anandan,  Phys. Rev. D,  {\bf 33, } No. 8, 2280-2287 (1986),

\bibitem{wi1961}
Eugene P. Wigner in {\it The Scientist Speculates}, I.J. Good, ed. (W. Heinemann, Ltd., London, 1961), reproduced in \cite{wi1970}, ch. 13.

\bibitem{pa1973}
J.C. Pati and Abdus Salam, Phys. Rev. {\bf 8,} 1240 (1973); Phys. Rev. Lett. {\bf 31,} 661 
(1973); Phys. Rev. {\bf D10} 275 (1974).  For a recent review, see J.C. Pati, 
hep-ph/0204240.

\bibitem{ge1974}
H. Georgi and S.L. Glashow, Phys. Rev. Lett. {\bf 32,} 438 (1974)



\end{thebibliography}
\end{document}